\documentclass[a4paper,aps,prd,10pt,preprintnumbers,showpacs,showkeys,twocolumn,superscriptaddress,nofootinbib,amsmath,amssymb]{revtex4-1}
\usepackage{cmap}
\usepackage[utf8]{inputenc}
\usepackage[T1]{fontenc}
\usepackage{graphicx,orcidlink,xcolor}
\def\re#1{Re(#1)}
\def\im#1{Im(#1)}

\def\Order#1{{\cal O}\left(#1\right)}
\def\imo{i}

\newcommand{\eq}[1]{\begin{align} #1 \end{align}}

\begin{document}

\title{Gravitational perturbations of Dymnikova black holes: grey-body factors and absorption cross-sections}

\author{Alexey Dubinsky}\email{dubinsky@ukr.net}
\affiliation{University of Seville, Seville, Spain}
%\pacs{04.30.-w,04.50.Kd,04.70.-s}

\begin{abstract}
We study axial gravitational perturbations of the Dymnikova regular black hole, an asymptotically flat spacetime in which the Schwarzschild singularity is replaced by a de Sitter core. Using the WKB method with Padé approximants, we compute grey-body factors, and absorption cross-sections, and test the recently proposed correspondence between quasinormal frequencies and transmission coefficients. We find that variations of the quantum parameter \(l_{\rm cr}\) affect the effective potential only near the horizon, leading to minor deviations of grey-body factors and absorption cross-sections from the Schwarzschild case. As a result, the Hawking radiation spectrum is governed mainly by the modified Hawking temperature, with grey-body factors providing only subleading corrections. Unlike higher quasinormal overtones, which are highly sensitive to near-horizon deformations, the grey-body factors remain robust, a feature explicitly confirmed for the Dymnikova geometry. The correspondence between quasinormal modes and grey-body factors holds in our case with high accuracy for multipoles $\ell \geq 2$.
\end{abstract}

\maketitle

\section{Introduction}

Quasinormal modes (QNMs) and grey-body factors (GBFs) constitute two key spectral signatures that capture complementary aspects of black-hole physics. Quasinormal modes characterize the damped oscillations of a black hole in response to external perturbations and dominate the ringdown stage of gravitational-wave signals \cite{Kokkotas:1999bd,Berti:2009kk,Konoplya:2011qq,Bolokhov:2025uxz}. Grey-body factors, on the other hand, quantify the frequency-dependent transmission probabilities for particles created via Hawking radiation \cite{Hawking:1975vcx} to escape through the curvature-induced potential barrier surrounding the horizon \cite{Page:1976df,Page:1976ki,Kanti:2004nr}. At first sight, these quantities are defined by distinct boundary conditions and appear unrelated: QNMs correspond to purely ingoing waves at the horizon and outgoing waves at infinity, whereas GBFs measure partial transmission of modes that originate near the horizon and scatter off the effective potential.  

Yet, a deeper connection has recently been uncovered. In \cite{Konoplya:2024lir} it was shown that, within the eikonal regime, the transmission coefficients of Hawking radiation can be expressed directly in terms of the fundamental quasinormal frequency
\[
\Gamma_{\ell}(\Omega) = \left[ 1 + \exp\!\left( \frac{2\pi\bigl(\Omega^{2}-\mathrm{Re}(\omega_{0})^{2}\bigr)}{4\,\mathrm{Re}(\omega_{0})\,\mathrm{Im}(\omega_{0})} \right) \right]^{-1}, \quad \ell \rightarrow \infty
\]
where $\omega_{0}$ denotes the fundamental quasinormal frequency.  Moreover, refinements of the WKB approximation allow this correspondence to extend beyond the strict eikonal limit, remaining accurate even for moderate values of the multipole number. This discovery is particularly significant because GBFs exhibit remarkable stability under small modifications of the effective potential, in contrast to higher quasinormal overtones which can vary drastically with perturbations of the geometry \cite{Konoplya:2022pbc,Shen:2025yiy}. Consequently, GBFs may provide a more robust diagnostic of the underlying spacetime, a property that has been emphasized in several recent analyses \cite{Rosato:2024arw,Rosato:2025byu,Wu:2024ldo,Konoplya:2025ixm}.  

The correspondence between QNMs and GBFs has already been generalized in multiple directions. It has been extended to the case of rotating black holes \cite{Konoplya:2024vuj} and even to nontrivial topologies such as traversable wormholes \cite{Bolokhov:2024otn}. Furthermore, it has been successfully tested and applied in a broad range of gravitational settings, demonstrating its versatility: effective quantum-corrected black holes \cite{Skvortsova:2024msa,Konoplya:2024lch,Tang:2025mkk}, the Bonanno–Reuter spacetime of asymptotically safe gravity \cite{Bolokhov:2025lnt}, Einstein–Gauss–Bonnet–Proca black holes with primary hair \cite{Lutfuoglu:2025ldc}, black holes embedded in dark-matter halos \cite{Hamil:2025pte}, Gibbons–Maeda–Garfinkle–Horowitz–Strominger black holes \cite{Dubinsky:2024vbn}, massive fields in Schwarzschild–de Sitter backgrounds \cite{Malik:2024cgb}, and higher-dimensional geometries \cite{Han:2025cal}.  

Among the various models of regular black holes, the solution proposed by Dymnikova~\cite{Dymnikova:1992ux} occupies a special place. It represents a static, spherically symmetric geometry in which the central singularity is replaced by a de Sitter core, while asymptotic flatness is preserved at large distances. This construction provides a simple yet consistent realization of singularity resolution within classical general relativity supplemented by a suitable effective matter source. More recently, the same functional form has reappeared in the context of Asymptotic Safety, where it arises naturally from renormalization group arguments fixing the running scale in terms of curvature invariants~\cite{Platania:2019kyx}. Thus, the Dymnikova spacetime serves both as one of the earliest examples of a regular black hole and as a concrete manifestation of quantum-gravity inspired corrections, making it a particularly interesting geometry to explore in detail.

The paper is organized as follows. In Section~II we review the Dymnikova geometry and its interpretation both as a phenomenological regular black hole and as a quantum-corrected solution arising in Asymptotic Safety. Section~III introduces the formalism of axial gravitational perturbations and derives the corresponding effective potential.  We also discuss there the boundary conditions for quasinormal modes and grey-body factors.  Section~IV presents numerical results for grey-body factors, and absorption cross-sections, and examines the accuracy of the correspondence between quasinormal frequencies and transmission coefficients.  Section V is devoted to remarks on the Hakwing radiation for Dymnikova black hole, while Section VI summarizes our findings and discusses their implications.

\section{The Dymnikova Metric}\label{sec:Dymnikova}

One of the earliest proposals for a regular black-hole geometry was given by Dymnikova in~\cite{Dymnikova:1992ux}. The central idea of this construction is to remove the curvature singularity of the Schwarzschild solution by replacing the central region with a de Sitter core while preserving asymptotic flatness at spatial infinity. This is achieved by considering a stress–energy distribution that violates some of the classical energy conditions but remains physically acceptable as an effective description of quantum backreaction. The resulting solution represents a static, spherically symmetric and asymptotically flat spacetime in which all curvature invariants remain finite everywhere.  

The Dymnikova line element takes the standard form
\begin{equation}\label{metric}
ds^2 = -f(r)\, dt^2 + \frac{dr^2}{f(r)} + r^2 \left(d\theta^2 + \sin^2\theta\, d\phi^2\right),
\end{equation}
with metric function
\begin{equation}
f(r) = 1 - \frac{2 M(r)}{r},
\end{equation}
where the mass function $M(r)$ interpolates smoothly between zero at the origin and the ADM mass $M$ at infinity. In its original form, Dymnikova proposed
\begin{equation}
M(r) = M \left(1 - e^{-r^3/r_0^3}\right),
\end{equation}
with $r_0$ setting the scale of the de Sitter core. At small radii the geometry approaches
\[
f(r) \approx 1 - \frac{r^2}{r_0^2},
\]
corresponding to a regular de Sitter interior, while for $r \gg r_0$ the standard Schwarzschild behavior $f(r) \approx 1 - 2M/r$ is recovered. Thus the spacetime is regular everywhere and smoothly interpolates between two familiar regimes.  

Although originally introduced as a phenomenological model, the same functional form was later recovered within the framework of Asymptotic Safety. In particular, Platania~\cite{Platania:2019kyx} demonstrated that a Dymnikova-type geometry naturally emerges once the classical Schwarzschild solution is modified by renormalization-group (RG) improvements of Newton’s constant.  

The starting point is the usual Schwarzschild lapse function,
\begin{equation}\label{lapse}
f(r) = 1 - \frac{2 M}{r},
\end{equation}
which solves the vacuum Einstein equations. The RG improvement consists of promoting the constant Newton coupling \(G_0\) to a running coupling \(G(r)\), depending on the radial scale. A widely used parametrization is
\begin{equation}\label{firstrepl}
G(r) = \frac{G_0}{1 + g_\ast^{-1} G_0\, k^2(r)},
\end{equation}
where \(k(r)\) is an effective cutoff scale and \(g_\ast\) denotes the dimensionless fixed-point value of the Newton coupling. The function \(k(r)\) is chosen such that \(k(r)\!\to\!0\) as \(r\!\to\!\infty\), ensuring the recovery of the classical Schwarzschild geometry at large distances. Substituting the running coupling into the metric yields
\begin{equation}
f(r) = 1 - \frac{2 M\, G(r)}{r G_0}.
\end{equation}

From the Einstein equations this modified lapse can be interpreted as arising from an effective energy–momentum tensor of the form
\begin{equation}\label{Teff}
T_{\mu\nu}^{\text{eff}} = (\rho+p)(l_\mu n_\nu + l_\nu n_\mu) + p g_{\mu\nu},
\end{equation}
where the null vectors satisfy \(l_\mu n^\mu=-1\). The effective energy density and pressure are generated by the radial variation of \(G(r)\):
\begin{equation}\label{qgfluid}
\rho = \frac{M G'(r)}{4 \pi r^2 G(r)}\,, \qquad 
p = -\frac{M G''(r)}{8 \pi r\, G(r)}.
\end{equation}
This stress tensor can be interpreted as encoding the vacuum polarization of the quantum gravitational field. In this picture, the energy density reflects the quantum self-energy associated with the running of Newton’s constant, and the feedback between geometry and coupling is implemented iteratively through the cutoff function \(k(r)\).  

The renormalization-group iteration proceeds by starting from the classical Schwarzschild background (\(G(r)=G_0\), \(T_{\mu\nu}^{\rm eff}=0\)), and successively updating \(G(r)\) via Eq.~\eqref{firstrepl}, with each step determining a new cutoff \(k(r)\) as a functional of the energy density generated at the previous step. In the continuum limit of infinitely many iterations one obtains a self-consistent solution in closed form. Remarkably, the resulting lapse function coincides with the Dymnikova metric,
\begin{equation}\label{f(r)}
f(r) = 1 - \frac{2 M}{r} \left(1 - e^{-r^3/(2 l_{\rm cr}^2 M)}\right).
\end{equation}
Here \(l_{\rm cr}\) is a critical length scale that governs the onset of quantum corrections. At $l_{cr}\to0$, the Schwarzschild spacetime is reproduced. For \(l_{\rm cr}\) below a maximum value,
\[
l_{\rm cr}^{\rm max} \simeq 1.138\, M,
\]
the function \eqref{f(r)} admits at least one event horizon. Above this threshold the horizon disappears, and the spacetime describes a horizonless, regular compact object.  

Thus, the Dymnikova geometry, first introduced phenomenologically in~\cite{Dymnikova:1992ux}, reappears as a natural outcome of Asymptotic Safety, where the de Sitter core arises dynamically from the running of Newton’s constant. This dual origin emphasizes its importance as a simple yet robust model of a singularity-free black hole. It is worth noting that the D-dimensional generalization of the Dymnikova metric may arise as a solution in theories with higher-curvature corrections \cite{Konoplya:2024kih}.

\section{Perturbation equations, effective potentials and boundary conditions}

The analysis of gravitational perturbations in this setting is considerably more subtle, since the background metric is obtained from an effective Hamiltonian-constraint approach rather than as an exact solution of Einstein’s equations with explicit quantum corrections. As a result, performing a fully rigorous treatment of perturbations is a difficult task. Nevertheless, as shown by Ashtekar, Olmedo, and Singh \cite{Ashtekar:2018lag,Ashtekar:2018cay}, quantum effects can be consistently mimicked by introducing an anisotropic fluid energy–momentum tensor within the Einsteinian framework. This reformulation makes it possible to analyze the dynamics of perturbations. Building on the approach of Bouhmadi-López et.al. \cite{Bouhmadi-Lopez:2020oia,Konoplya:2024lch}, one can study axial perturbations under the simplifying assumption that fluctuations along the anisotropy direction do not contribute in the axial sector.  

This strategy mirrors the analysis of \cite{Bronnikov:2012ch}, where black holes coupled to scalar fields were considered and scalar perturbations decoupled from axial gravitational modes (see also \cite{Chen:2019iuo}). Such analogies strengthen confidence in the applicability of this approach, even if some sector-specific features may remain inaccessible.

After separation of variable and some algebra, introducing the new variables makes
the perturbation equation take the Schrödinger-like form
\eq{
\frac{d^2}{dr_{*}{}^2}\Psi+\left(\omega^2-V(r)\right)\Psi=0,
}
with the effective axial potential
\eq{
V=f(r)\left(\frac{2f(r)}{r^2}-\frac{f'(r)}{r}+\frac{(\ell+2)(\ell-1)}{r^2}
\right).
}
Here, the tortoise coordinate is $dr^*=dr/f(r)$ and $\ell=2,3,4,\ldots$ is the multipole number arising from separation of variables. It should be emphasized that this effective potential does not coincide with the Regge–Wheeler potential of the vacuum Schwarzschild solution, since the perturbations also affect the energy–momentum tensor of the fluid.  

\begin{figure*}
\resizebox{\linewidth}{!}{\includegraphics{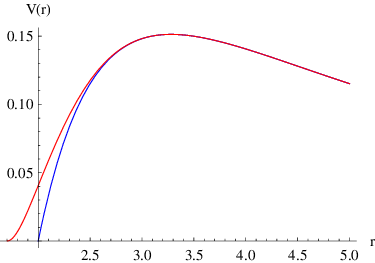}\includegraphics{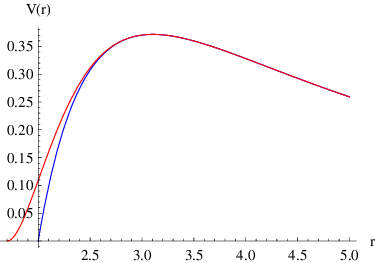}}
\caption{Effective potentials for $\ell=2$ (left) and $\ell=3$ (right) axial gravitational perturbations: $M=1$, $l_{\rm cr}=0.01$ (blue, bottom), $l_{\rm cr}=1.138$ (red, top).}\label{fig:Pot}
\end{figure*}

Figure \ref{fig:Pot} illustrates the effective potentials for axial gravitational perturbations of the Dymnikova black hole. For very small values of the parameter \(l_{\rm cr}\) (e.g., \(l_{\rm cr}=0.01\)  units of black-hole mass), the geometry approaches the Schwarzschild limit, and the potential differs from the Schwarzschild case only in the near-horizon region. The deformation is localized close to the horizon and rapidly merges with the Schwarzschild profile at larger radii. At the same time, the position of the event horizon shifts inward, indicating that the effective horizon radius decreases as $l$ increases from zero. This behavior highlights the fact that quantum-inspired corrections encoded in the Dymnikova geometry predominantly affect the near-horizon structure while leaving the asymptotic region essentially unchanged.

Similar simplifications have been employed in a number of related contexts, where certain perturbations are assumed negligible and consequently omitted \cite{Berti:2003yr,Kokkotas:1993ef,Konoplya:2006ar}. While this procedure may omit potentially interesting corrections to the full gravitational spectrum, it nonetheless provides a good first approximation when deviations from the Schwarzschild geometry remain small. This viewpoint is consistent with the perturbative nature of quantum corrections, which are expected to affect the spacetime only modestly.  

For asymptotically flat black holes, the quasinormal modes (QNMs) are defined by wave-like solutions of the perturbation equation subject to purely radiative boundary conditions. Near the event horizon, physical waves can only fall inward, so the perturbation behaves as an ingoing mode in the tortoise coordinate $r_\ast \to -\infty$. At spatial infinity, by contrast, no radiation is allowed to come in from outside, and the solution must be purely outgoing as $r_\ast \to +\infty$. These conditions select a discrete set of complex frequencies $\omega = \omega_{\text{Re}} - i \omega_{\text{Im}}$, where $\omega_{\text{Re}}$ determines the oscillation frequency and $\omega_{\text{Im}}>0$ governs the damping rate. The QNM spectrum therefore captures the intrinsic oscillatory response of the black hole and determines the ringdown phase of gravitational-wave signals.  

\begin{equation}
\Psi(r_\ast) \propto
\begin{cases}
e^{-i\omega r_\ast}, & r_\ast \to -\infty \quad (\text{horizon}), \\[0.2cm]
e^{+i\omega r_\ast}, & r_\ast \to +\infty \quad (\text{infinity}),
\end{cases}
\end{equation}

The grey-body factors (GBFs) \cite{Page:1976df,Page:1976ki,Kanti:2004nr}., in turn, are defined through a scattering problem rather than a resonance condition. In this case, one considers a unit flux of ingoing waves at the horizon, which partially transmits through the potential barrier and partially reflects back. At spatial infinity the solution is a superposition of outgoing and reflected ingoing waves, and the transmission coefficient $\Gamma_\ell(\Omega)$ is extracted from the relative amplitudes of these components. Unlike QNMs, which form a discrete spectrum, the GBFs are smooth functions of the {\it real frequency $\Omega$} and encode how the effective potential surrounding the black hole filters Hawking radiation. Together, QNMs and GBFs provide complementary characterizations of black-hole spacetimes: the former probe resonant oscillations, while the latter quantify scattering and particle emission.  
\begin{widetext}
\begin{equation}
\Psi(r_\ast) = 
\begin{cases}
e^{-i\Omega r_\ast}, & r_\ast \to -\infty \quad (\text{unit ingoing flux at horizon}), \\[0.2cm]
A_{\rm out}\, e^{+i\Omega r_\ast} + A_{\rm in}\, e^{-i\Omega r_\ast}, & r_\ast \to +\infty \quad (\text{scattering at infinity}),
\end{cases}
\end{equation}
\end{widetext}
with transmission coefficient (grey-body factor)
\[
\Gamma_\ell(\Omega) = 1- \frac{|A_{\rm out}|^2}{|A_{\rm in}|^2 }\, .
\]

\begin{figure*}
\resizebox{\linewidth}{!}{\includegraphics{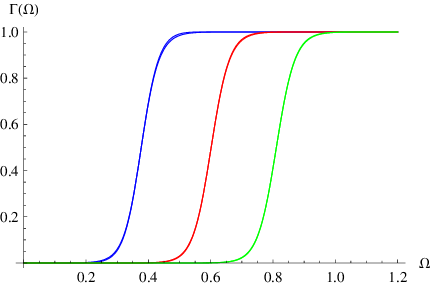}\includegraphics{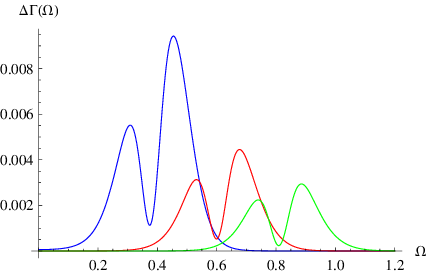}}
\caption{Grey-body factors found by the 6th order WKB formula (left) and via the correspondence (right) \textbf{difference?} for $M=1$, $l_{\rm cr}=0.01$.}\label{fig:Gamma}
\end{figure*}

\begin{figure*}
\resizebox{\linewidth}{!}{\includegraphics{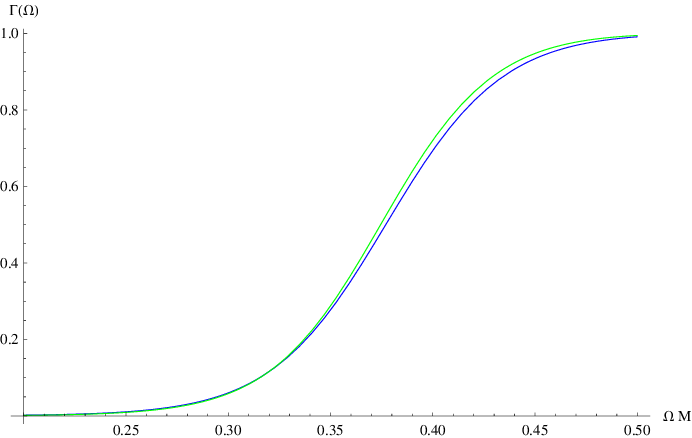}\includegraphics{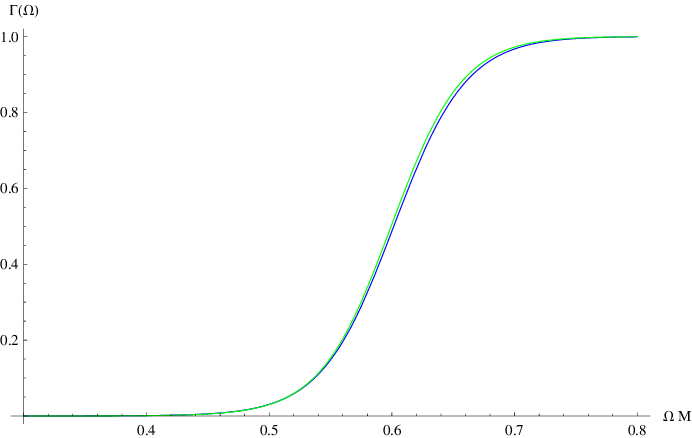}}
\caption{Grey-body factors found via the correspondence for $M=1$, $l_{\rm cr}=0.01$ (blue) and $l_{\rm cr}=1.138$ (green) for $\ell=2$ (left) and $\ell=3$ (right).}\label{fig:Gamma2}
\end{figure*}

\begin{figure*}
\resizebox{\linewidth}{!}{\includegraphics{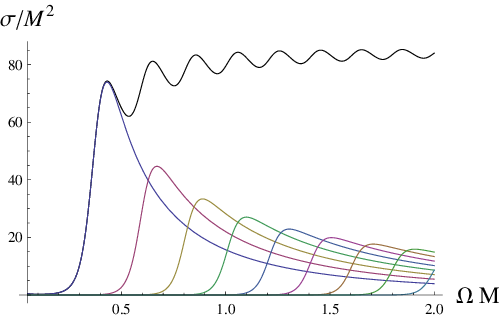}\includegraphics{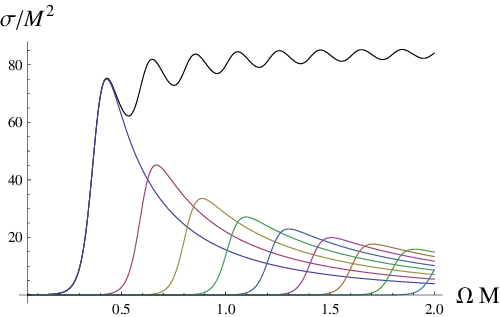}}
\caption{Absorption cross-sections for $M=1$, $l_{\rm cr}=0.01$ (left), $l_{\rm cr}=1.138$ (right).}\label{fig:SIGMA}
\end{figure*}

\section{Grey-Body Factors and Absorption Cross-Section}

While direct numerical integration of the perturbation equation yields the most accurate values of the transmission coefficients $\Gamma_{\ell}(\Omega)$ (see seminal works of Don Page \cite{Page:1976df,Page:1976ki}), it is often preferable to rely on semi-analytical methods, which are computationally faster and provide useful physical insight. Among these approaches, the Wentzel–Kramers–Brillouin (WKB) technique has become the standard tool for black-hole scattering problems. Initially applied to quasinormal-mode spectra in~\cite{Schutz:1985km,Iyer:1986np,Konoplya:2003ii}, the method is based on connecting asymptotic solutions across the classically forbidden region around the peak of the potential barrier, thereby producing approximate expressions for reflection and transmission.  

In practice, the WKB approximation is formulated as an expansion around the maximum of the effective potential. At leading order, the grey-body factor takes the form \cite{Schutz:1985km}  
\begin{equation}
\Gamma_{\ell}(\Omega) = \frac{1}{1+\exp\!\left(2\pi i K\right)},
\end{equation}
with $K$ determined by the barrier height $V_{0}$ and its second derivative $V_{0}''$ at the peak. Systematic higher-order corrections incorporate progressively higher derivatives of the potential \cite{Matyjasek:2017psv,Konoplya:2019hlu}, thereby improving the accuracy of the approximation. In this work we employ the sixth-order expansion developed in~\cite{Konoplya:2003ii}, which has proven reliable across a wide class of backgrounds.  The WKB–based framework has been extensively applied in contexts ranging from classical general relativity to theories with higher-curvature or quantum corrections. Various developments and applications of this approach can be found in~\cite{Pedrotti:2025idg,Antonelli:2025yol,Konoplya:2005sy,Lutfuoglu:2025ljm,Zhang:2025xqt,Konoplya:2020cbv,Konoplya:2021ube,Matyjasek:2021xfg,Miyachi:2025ptm,Matyjasek:2017psv,Konoplya:2023moy,Hamil:2025cms,MahdavianYekta:2019pol}, which illustrate the versatility of the method in diverse gravitational settings.

The higher-order WKB method, improved by Padé resummation, provides a powerful semi-analytical tool for calculating quasinormal modes of black holes. In this approach, the wave equation is expanded near the peak of the effective potential, and the WKB quantization condition relates the complex frequency $\omega$ to the potential and its derivatives. For an $N$-th order expansion the condition takes the form  
\begin{equation}
\frac{i(\omega^2 - V_0)}{\sqrt{-2 V_0''}} - \sum_{j=2}^{N} \Lambda_j = n + \tfrac{1}{2},
\end{equation}
where $V_0$ and $V_0''$ are the value and second derivative of the effective potential at its maximum, $\Lambda_j$ are higher-order correction terms, and $n=0,1,2,\ldots$ is the overtone number. Padé approximants of the WKB series are then constructed to accelerate convergence and significantly improve accuracy, allowing for reliable extraction of QNMs even for moderate multipole numbers. To find the two low lying quasinormal modes $\omega_0$ and $\omega_1$, we used the sixth-order WKB method with Padé approximants $\tilde{m} = \tilde{n} = 3$,  \cite{Konoplya:2019hlu}, which results in the best accuracy in numerous cases \cite{Malik:2023bxc,Skvortsova:2023zmj,Bolokhov:2023bwm,Dubinsky:2025fwv,Dubinsky:2025azv,Dubinsky:2024rvf,Konoplya:2020jgt,Skvortsova:2024atk,Lutfuoglu:2025hjy,Bronnikov:2019sbx,Churilova:2021tgn,Malik:2024nhy,Dubinsky:2024nzo,Dubinsky:2025bvf,Konoplya:2019xmn,Dubinsky:2024hmn,Skvortsova:2024wly,Bolokhov:2024ixe}

In \cite{Konoplya:2024lir}, it was found that $K$ can be expressed through the fundamental mode $\omega_0$ and the first overtone $\omega_1$ for any fixed $\ell$ and the explicit expression depends only on those frequencies, but not on $\ell$, 
\begin{eqnarray}\nonumber
&&\imo K=\frac{\Omega^2-\re{\omega_0}^2}{4\re{\omega_0}\im{\omega_0}}\Biggl(1+\frac{(\re{\omega_0}-\re{\omega_1})^2}{32\im{\omega_0}^2}
\\\nonumber&&\qquad\qquad-\frac{3\im{\omega_0}-\im{\omega_1}}{24\im{\omega_0}}\Biggr)
-\frac{\re{\omega_0}-\re{\omega_1}}{16\im{\omega_0}}
\\\nonumber&& -\frac{(\Omega^2-\re{\omega_0}^2)^2}{16\re{\omega_0}^3\im{\omega_0}}\left(1+\frac{\re{\omega_0}(\re{\omega_0}-\re{\omega_1})}{4\im{\omega_0}^2}\right)
\\\nonumber&& +\frac{(\Omega^2-\re{\omega_0}^2)^3}{32\re{\omega_0}^5\im{\omega_0}}\Biggl(1+\frac{\re{\omega_0}(\re{\omega_0}-\re{\omega_1})}{4\im{\omega_0}^2}
\\\nonumber&&\qquad +\re{\omega_0}^2\Biggl(\frac{(\re{\omega_0}-\re{\omega_1})^2}{16\im{\omega_0}^4}
\\&&\qquad\qquad -\frac{3\im{\omega_0}-\im{\omega_1}}{12\im{\omega_0}}\Biggr)\Biggr)+ \Order{\frac{1}{\ell^3}}.
\label{eq:gbsecondorder}
\end{eqnarray}

\begin{figure*}
\resizebox{\linewidth}{!}{\includegraphics{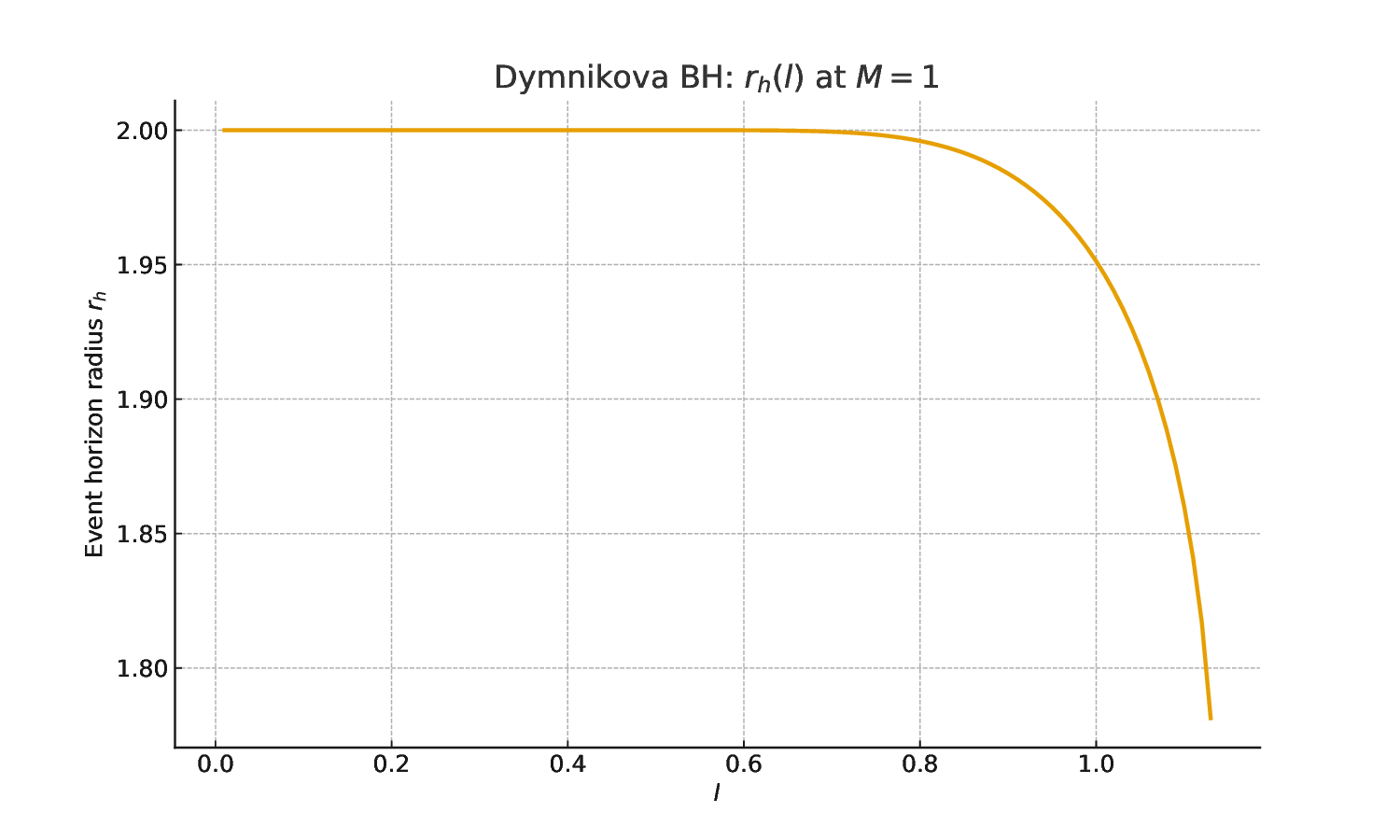}\includegraphics{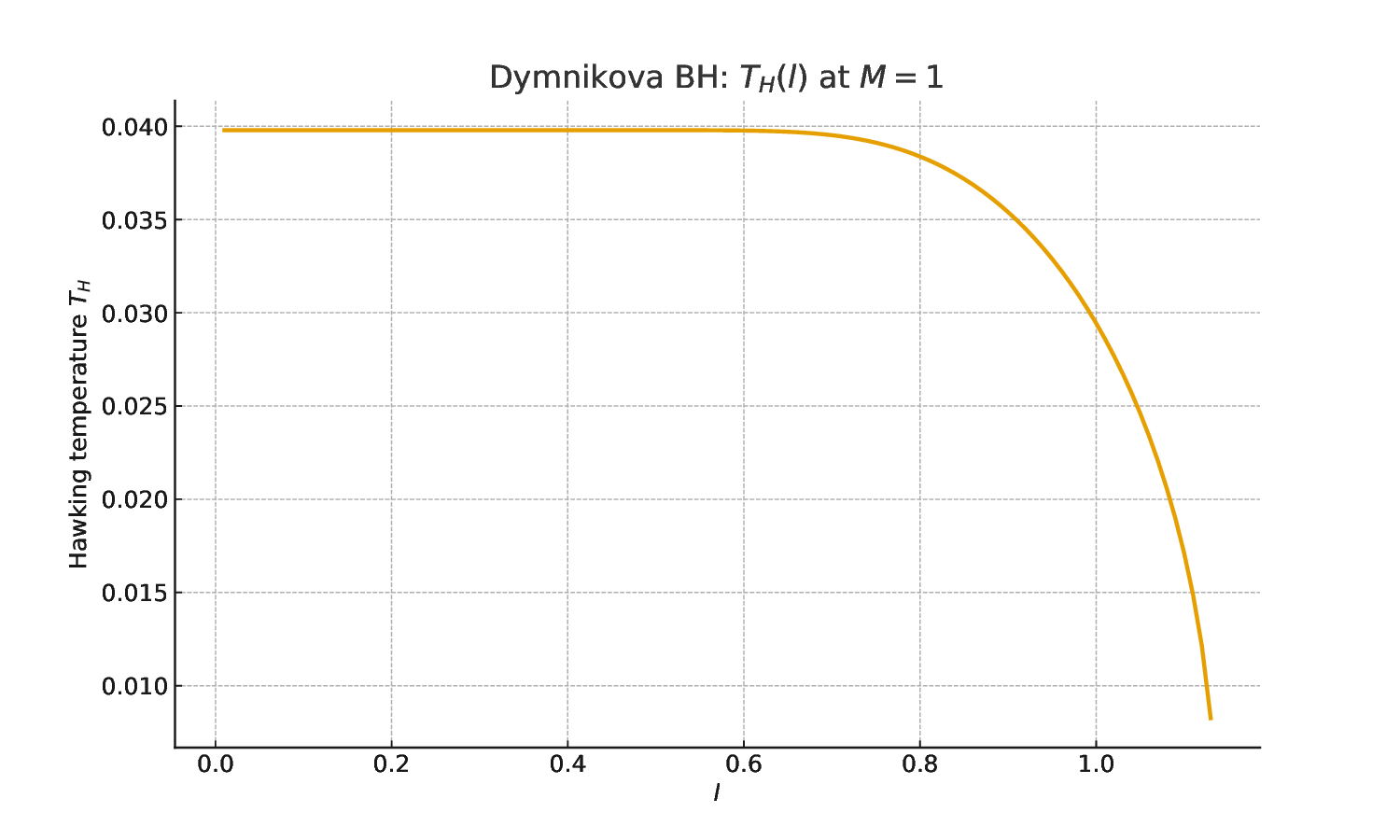}}
\caption{The radius of the event horizon (left) and Hawking temeprature (right) as a function of the parameter \(l_{\rm cr}\) at \(M=1\).}\label{fig:Temperature}
\end{figure*}

In our case, one observes that as the quantum parameter \(l_{\rm cr}\) varies from zero to its near-extremal value, the grey-body factors are only weakly affected (see fig. \ref{fig:Gamma2}). This is because the effective potential near its maximum rapidly approaches the Schwarzschild form, leading to only minor deviations. The relative error of the correspondence remains within a few percent for $\ell=2$ and decreases rapidly with increasing multipole number. Although this provides further confirmation of the correspondence in the present case, as well as in a number of other recent studies \cite{Lutfuoglu:2025ldc,Bolokhov:2025lnt,Hamil:2025pte,Konoplya:2024lch,Han:2025cal,Lutfuoglu:2025ohb,Lutfuoglu:2025hjy,Malik:2025dxn,Malik:2024cgb,Dubinsky:2024vbn,Shi:2025gst}, it cannot be taken as evidence that the correspondence between grey-body factors and quasinormal modes is universal.  

Indeed, the correspondence may work poorly, or even fail completely, when the effective potential develops multiple peaks or when additional branches of modes appear that are inaccessible to the WKB method. In this respect, the situation parallels the correspondence between null geodesics and eikonal quasinormal modes, which is also known to break down in various higher-curvature theories \cite{Konoplya:2020bxa,Konoplya:2025afm,Bolokhov:2023dxq,Konoplya:2017wot} and in asymptotically de Sitter spacetimes \cite{Konoplya:2022gjp,Konoplya:2025mvj}. In the latter case, the failure arises because the spectrum contains an additional branch of modes, those of pure de Sitter space, that cannot be captured by the WKB approximation.

The grey-body factors at different multipole numbers can be combined to extract another key quantity characterizing the interaction of radiation with a black hole: the absorption cross-section~\cite{Futterman:1988ni},  
\begin{equation}
\sigma(\Omega) = \frac{\pi}{\Omega^{2}} \sum_{\ell=2}^{\infty} (2\ell+1)\,\Gamma_{\ell}(\Omega).
\end{equation}
This cross-section $\sigma(\Omega)$ plays the role of an effective geometric area that the black hole presents to an incoming wave of frequency $\Omega$. From a physical perspective, it measures the likelihood that incident radiation penetrates the potential barrier and is absorbed by the black hole, instead of being scattered back to infinity. Recent analyses of this quantity in a variety of black-hole settings can be found in~\cite{Li:2024xyu,Heidari:2024bkm,Polo:2024xqm,OuldElHadj:2025hbl,C:2024cnk}.  

As shown in Fig.~\ref{fig:SIGMA}, the absorption cross-section is only mildly sensitive to the quantum parameter \(l_{\rm cr}\) in the low-frequency regime. The inclusion of the correction slightly shifts the curve but does not lead to qualitative changes, reflecting once again that quantum effects in the Dymnikova background are predominantly confined to the near-horizon region.

\section{Remarks on Hawking radiation}

Although there are cases where grey-body factors play an important role in determining the intensity of Hawking evaporation \cite{Konoplya:2019ppy}, in most situations the Hawking temperature remains the dominant factor.  In the previous section we observed that the grey-body factors and absorption cross-section also do not change when the quantum parameter is changed from the Schwarzschild limit to the extreme quantum corrected black hole. Therefore, an approxiamte expression for the intensity of Hawking radiation will include the grey-body factors for the Schwarzschild black hole and Hawking temperature of Dymanikova BH.

The energy emission rate per unit frequency $\Omega$ is generally given by \cite{Hawking:1975vcx}
\begin{equation}
\frac{d^{2}E}{d\Omega\,dt}
= \frac{1}{2\pi}\;\sum_{\ell=0}^{\infty} (2\ell+1)\,
\frac{\Gamma_\ell(\Omega)}{e^{\Omega/T_H}-1}\,\Omega ,
\end{equation}
where $\Gamma_\ell(\Omega)$ are the grey-body factors and $T_H$ is the Hawking temperature.

Introducing the absorption cross-section
\begin{equation}
\sigma_{\rm abs}(\Omega)=\frac{\pi}{\Omega^{2}}\sum_{\ell=0}^{\infty}(2\ell+1)\,\Gamma_\ell(\Omega),
\end{equation}
the spectrum can be written in the equivalent form
\begin{equation}
\frac{d^{2}E}{d\Omega\,dt}
= \frac{\sigma_{\rm abs}(\Omega)}{2\pi^{2}}\,
\frac{\Omega^{3}}{e^{\Omega/T_H}-1}\, .
\end{equation}
Since the absorption cross-section is very close to that for the Schwarzschild black hole, we can deduce the approximate formula for the intensity of Hawking radiation 
\begin{equation}
\frac{d^{2}E}{d\Omega\,dt}
\approx \frac{\sigma_{\rm abs}^{Schw}(\Omega)}{2\pi^{2}}\,
\frac{\Omega^{3}}{e^{4 \pi \Omega/f'(r_{H})}-1}\, ,
\end{equation}
where $\sigma_{\rm abs}^{Schw}(\Omega)$ is the cross-section of the Schwarzschild black hole.
Since the Hawking temperature decreases significantly as the quantum parameter \(l_{\rm cr}\) increases (see Fig.~\ref{fig:Temperature}), the energy emission rate of the quantum-corrected black hole is correspondingly strongly suppressed.  

In this work we have restricted our analysis to axial perturbations; although polar perturbations may in principle lead to different spectra, the fact that the Dymnikova geometry deviates from Schwarzschild only in the near-horizon region suggests that the grey-body factors will likewise differ only slightly, so that multiplicity factors could safely be taken as $2 (2\ell+1)$ for all types of gravitons.

%\vspace{5mm}
\section{Conclusions}

We have analyzed axial gravitational perturbations of the Dymnikova black hole and computed the associated quasinormal modes, grey-body factors, and absorption cross-sections. Our results show that the quantum parameter \(l_{\rm cr}\), which regularizes the central singularity, affects the effective potential only in the near-horizon region. As a consequence, the grey-body factors deviate only marginally from those of Schwarzschild, while the absorption cross-sections exhibit only small shifts in the low-frequency regime. The dominant impact of $l$ on Hawking radiation arises instead through the modified Hawking temperature, which decreases toward the near-extremal limit.  

We have also confirmed that the approximate correspondence between quasinormal modes and grey-body factors holds with good accuracy for the Dymnikova spacetime, with relative errors of only a few percent for $\ell=2$ and rapidly diminishing at higher multipoles. At the same time, as emphasized in earlier works, this relation is not universal and may fail in situations where the effective potential acquires multiple peaks or when additional branches of modes, inaccessible to the WKB approximation, appear in the spectrum.  

An interesting aspect of our results concerns the relative sensitivity of quasinormal spectra and grey-body factors to near-horizon modifications of the geometry. It was shown in \cite{Konoplya:2025ixm} that while the higher overtones of quasinormal modes can react very strongly to even small deformations of the effective potential close to the horizon \cite{Konoplya:2022pbc,Konoplya:2023aph}, the grey-body factors remain comparatively stable. The Dymnikova black hole provides a concrete realization of this general observation: despite the fact that the near-horizon region is altered by the parameter \(l_{\rm cr}\), the transmission coefficients and absorption cross-sections deviate only slightly from their Schwarzschild values. Thus, grey-body factors appear to be considerably more robust indicators of the spacetime geometry than the overtone sector of the quasinormal spectrum.

Overall, our study demonstrates that the Dymnikova black hole, despite its regular core, radiates in a manner that is nearly indistinguishable from Schwarzschild at the level of grey-body factors and absorption cross-sections, with the primary modification being the shift of the Hawking temperature. This reinforces the robustness of Schwarzschild-like emission properties in a wide class of regular and quantum-corrected black-hole geometries.

\begin{acknowledgments}
The author thanks R. A. Konoplya for useful discussions. The author acknowledges the University of Seville for their support through the Plan-US of aid to Ukraine.
\end{acknowledgments}

\bibliography{bibliography}

\end{document}